\documentclass[preprint,portrait,dvips,notoc]{JHEP3}
\usepackage{graphicx}
\usepackage{amsmath}
\usepackage{psfrag}
\usepackage{slashed}
\usepackage[sort&compress,numbers]{natbib}

\def\la{\langle}
\def\ra{\rangle}

\def\A#1#2{\la#1#2\ra}
\def\B#1#2{[#1#2]}
\def\AB#1#2#3{\la#1|#2|#3]}

\def\spa#1.#2{\left\langle#1\,#2\right\rangle}
\def\spb#1.#2{\left[#1\,#2\right]}
\def\spab#1.#2.#3{\left\langle#1\,#2\,#3\right]}

\def\spa#1.#2{\left\langle#1\,#2\right\rangle}
\def\spb#1.#2{\left[#1\,#2\right]}
\def\spab#1.#2.#3{\left\langle#1\,#2\,#3\right]}
\def\spba#1.#2.#3{\left[#1\,#2\,#3\right\rangle}
\def\spaa#1.#2.#3.#4.#5.#6{\left\langle#1\,#2\,#3\,#4\.#5\,#6\right\rangle}
\def\spbb#1.#2.#3.#4.#5.#6{\left [#1\,#2\,#3\,#4\,#5\,#6\right ]}
\def\spbnin#1.#2.#3.#4.#5.#6.#7.#8.#9{\left [#1|\,#2\,#3\,#4\,#5\,#6\,#7\,#8|\,#9\right \rangle}
\def\spanin#1.#2.#3.#4.#5.#6.#7.#8.#9{\left \langle#1|\,#2\,#3\,#4\,#5\,#6\,#7\,#8|\,#9\right]}
\def\spasev#1.#2.#3.#4.#5.#6.#7{\left \langle#1|\,#2\,#3\,#4\,#5\,#6\,|#7\right]}
\def\spafiv#1.#2.#3.#4.#5{\left \langle#1|\,#2\,#3\,#4\,|#5\right]}
\def\spbsev#1.#2.#3.#4.#5.#6.#7{\left [#1|\,#2\,#3\,#4\,#5\,#6|\,#7\right \rangle}
\def\spbsix#1.#2.#3.#4.#5.#6{\left [#1|\,#2\,#3\,#4\,#5|\,#6\right ]}
\def\spbfiv#1.#2.#3.#4.#5{\left [#1|\,#2\,#3\,#4|\,#5\right \rangle}
\def\spbf#1.#2.#3.#4{\left [#1\,#2\,#3\,#4\right ]}
\def\spahr#1.#2{\langle#1\,\hat{#2}\rangle}
\def\spaah#1.#2.#3.#4{\langle#1\,#2\,#3\,\hat{#4}\rangle}
\def\spaahl#1.#2.#3.#4{\langle\hat{#1}\,#2\,#3\,#4\rangle}
\def\spabh#1.#2.#3{\langle#1\,\widehat{#2}\,#3]}
\def\spahl#1.#2{\langle\hat{#1}\,#2\rangle}
\def\spahh#1.#2{\langle\hat{#1}\,\hat{#2}\rangle}
\def\spaas#1.#2.#3{\left\langle#1\,#2\,#3\right\rangle}
\def\spbhl#1.#2{\left[\hat{#1}\,#2\right]}
\def\spbhr#1.#2{\left[#1\,\hat{#2}\right]}
\def\spabhh#1.#2.#3{\langle\hat{#1}\,\hat{#2}\,#3]}
\def\spbah#1.#2.#3{[#1\,\widehat{#2}\,#3\rangle}
\def\mc#1{\mathcal{#1}}

\DeclareMathOperator{\FFtme}{ {\rm F}^{2me}_{4F}}
\DeclareMathOperator{\FFtmh}{ {\rm F}^{2mh}_{4F}}

\DeclareMathOperator{\FFom}{ {\rm F}^{1m}_{4F}}
\DeclareMathOperator{\Ftmtri}{{\rm F}^{3m}_3}

\DeclareMathOperator{\tr}{\rm tr}

\def\e{\epsilon}
\def\qb{{\bar{q}}}

\def\kf#1{K_{#1}^\flat}
\def\kfm#1{K_{#1}^{\flat\mu}}

\preprint{IPPP/09/58 \\
CERN-PH-TH/2009-163 \\
DESY-09-138 \\
\today}

\title{One-loop Higgs plus four gluon amplitudes: Full analytic results}

\author{Simon Badger$^\dag$,
    \ E. W. Nigel Glover$^*$,
    \ Pierpaolo Mastrolia$^\ddag$,
    \ Ciaran Williams$^*$
    	\\
	$^\dag$Deutches Elektronen-Synchrotron DESY, Platanenallee, 6, D-15738 Zeuthen, Germany
	\\
	$^*$Department of Physics, University of Durham, Durham, DH1 3LE, UK
	\\
        $^\ddag$Theory Division, CERN CH-1211 Geneva 23, Switzerland
	\\
	E-mails: 
	{\tt simon.badger@desy.de},
        {\tt e.w.n.glover@durham.ac.uk}, 
        {\tt pierpaolo.mastrolia@cern.ch}, 
	{\tt ciaran.williams@durham.ac.uk}.
}

\abstract{
We consider one-loop amplitudes of a Higgs boson coupled to gluons in the limit of a large top
quark mass. We treat the Higgs as the real part of a complex field $\phi$ that couples to the self-dual
field strengths and compute the one-loop corrections to the $\phi$-NMHV amplitude, which
contains one gluon of positive helicity whilst the remaining three have negative helicity.  We use
four-dimensional unitarity to construct the cut-containing contributions and a hybrid of Feynman
diagram and recursive based techniques to determine the rational piece. Knowledge of the
$\phi$-NMHV contribution completes the analytic  calculation of the Higgs plus four gluon amplitude.
For completeness we also include expressions for the remaining helicity configurations which have
been calculated elsewhere. These amplitudes are relevant for Higgs plus jet production via gluon
fusion in the limit where the top quark is large compared to all other scales in the problem.
\\\today
}


\begin{document}

{\allowdisplaybreaks
\section{Introduction}

The search for the Higgs boson is one of the primary objectives of the Large Hadron Collider (LHC).  If
discovered, attention will swiftly turn to the exploration of the Higgs sector and the measurement of the
Higgs couplings to the weak gauge bosons and to fermions. The main Higgs production processes at the LHC
are gluon fusion, which proceeds via a top quark loop, and vector boson fusion (VBF), which is dominated by
the $t$-channel exchange of weak bosons.   The Higgs production rate via vector boson fusion is typically
about an order of magnitude smaller than the gluon fusion rate, but has a characteristic signature with two
forward quark jets.      The next-to-leading order (NLO) QCD corrections to the VBF process are small
(5-10\%) \cite{Han:1992hr,Figy:2003nv,Figy:2004pt,Berger:2004pca} which makes vector boson fusion very
attractive for the measurement of the Higgs coupling to weak bosons.   Recently, the full EW+QCD
corrections to this process have been computed~\cite{Ciccolini:2007jr,Ciccolini:2007ec}.

The dominant background to VBF comes from Higgs plus two jet production via gluon fusion.  Leading order
calculations have been performed both with the exact $m_t$ dependence \cite{DelDuca:H2j1} and in the large
$m_t$ limit~\cite{Dawson:Htomultijet,Kauffman:H2jets} where the top quark loop is replaced by an effective
local interaction $C(m_t) H
G^{\mu\nu}G_{\mu\nu}$~\cite{Wilczek:HggOperator5,Djouadi:NLOHiggs1,Dawson:NLOHiggs2}.  For inclusive Higgs 
production it has been shown that this approximation is valid over a large range of Higgs masses
\cite{Kramer:1996iq}.  The approximation remains valid for processes with increased numbers of jets
provided that the transverse momenta of the jets is smaller than $m_t$~\cite{DelDuca:H2j3}.  

The next-to-leading order QCD corrections to the gluon fusion Higgs plus two jet rate have been calculated
numerically in the large-$m_t$ limit~\cite{Campbell:NLOHjj} using the real radiative corrections of
Ref.~\cite{DelDuca:Hto3jets,Dixon:HiggsMHV,Badger:HpartonMHV} and the partly analytic, partly numerical virtual one-loop corrections of
Ref.~\cite{Ellis:1lh24}.  There are four distinct processes that contribute to the virtual NLO corrections,
$gg\to ggH$, $gg \to q\bar q H$, $q\bar q \to q  \bar q  H$, $q\bar q \to q^\prime \bar q^\prime H$ and the
associated crossings. Ref.~\cite{Ellis:1lh24} provided analytic expressions for the spin- and colour-averaged one-loop ``squared" matrix elements for
$q\bar q \to q  \bar q  H$ and  $q\bar q \to q^\prime \bar q^\prime H$ and evaluated the other two
processes numerically.   Because of the size of the NLO QCD corrections, for studies of Higgs phenomenology
it is important to incorporate the NLO QCD corrections to Higgs plus two jets in an efficient and flexible
way. This has led to a collective effort to derive compact analytic expressions for the one-loop
corrections for the $gg\to ggH$ and $gg \to q\bar q H$ processes.

An effective way of deriving compact analytic expressions for the Higgs plus four parton amplitudes is to employ on-shell unitarity methods. 
The original unitarity method involved using four-dimensional double cuts of
amplitudes to classify the coefficients of discontinuities associated with
physical invariants \cite{BDDK:uni1,BDDK:uni2}. Unitarity cuts found many
applications in the calculations of amplitudes in $\mc{N}=4$ and $\mc{N}=1$
supersymmetric Yang-Mills theories, since these amplitudes can be fully
constructed from their unitarity cuts. Amplitudes in non-supersymmetric theories
contain rational terms which cannot be determined from four dimensional cuts but
can be computed using a number of complementary methods~\cite{Bern:1995db,Bern:2005hs,Binoth:2006hk,Xiao:rationalI,Su:rationalII,Xiao:rationalIII}.  

In recent years the unitarity method has been generalised to include multiple cuts
\cite{Britto:2004nc,Forde:intcoeffs,Mastrolia:2006ki,Britto:sqcd,Britto:ccqcd,BjerrumBohr:2007vu,Glover:2008ur,Ossola:2008xq},
this has lead to an explosion in the number of unitarity based techniques in calculations of NLO processes.
In particular by working in $D$-dimensions
\cite{Bern:1995db,Bern:SDYM,Anastasiou:DuniII,Anastasiou:DuniI,Badger:rational}, one can completely
determine the amplitude from its unitarity cuts. This is because the rational parts of one-loop
amplitudes arise from higher-dimensional basis integrals, which are sensitive to $D$-dimensional cuts.  At
tree-level the discovery of another on-shell method, the BCFW recursion relations
\cite{Britto:rec,Britto:proof}, also sparked developments at one-loop. Recognising that the rational part
of one-loop amplitudes also satisfied recursion relations allowed a fully four dimensional on-shell method
to be introduced (the unitarity bootstrap) \cite{Bern:2005hs,Berger:genhels}. Several groups have
produced sophisticated numerical programs based on on-shell methods which aim to efficiently calculate NLO
one-loop amplitudes of relevance to the LHC \cite{Ossola:2007ax,Ellis:2008kd,Berger:2008sj}.  

In order to efficiently compute one-loop amplitudes using on-shell unitarity based techniques, it is
desirable to have compact tree amplitudes. For amplitudes involving a Higgs boson coupling to partons, it is
convenient to split the real Higgs scalar into two complex scalars ($\phi$ and $\phi^{\dagger}$) such that 
$H = \phi+\phi^{\dagger}$~\cite{Dixon:HiggsMHV}.  In this case, the effective Higgs-gluon interaction also
separates into two parts - the $\phi$ couples directly to the self-dual gluon field strengths, whilst
$\phi^{\dagger}$ couples to the anti-self-dual gluon field strengths. The tree $\phi$-amplitudes are
MHV-like and have a very simple structure. $\phi^\dagger$ amplitudes are obtained from $\phi$
amplitudes by complex conjugation. With this breakdown, amplitudes for a light pseudo-scalar Higgs boson
$A$ can be constructed from the difference of $\phi$- and $\phi^\dagger$-amplitudes,  $A \sim
\phi-\phi^{\dagger}$. The $A$-amplitudes may be relevant amplitudes for SUSY theories, provided that $m_{A}
< 2 m_{t}$.  

Analytic calculations of one-loop $\phi$-amplitudes to date have not included NMHV helicity configurations.
Previous one-loop calculations for amplitudes with $\phi$ coupling to any number of gluons have restricted
the number of negative helicity gluons to: none or one (which are rational
amplitudes)~\cite{Berger:phinite}, two (the $\phi$-MHV 
configurations)~\cite{Badger:1lphiMHV,Glover:1lphiMHVgeneral}. An expression for the amplitude containing
only negative helicity gluons has also been obtained~\cite{Badger:1lhiggsallm}. Recently new analytic
results for the helicity amplitudes of the $\phi q\overline{q}g^{\pm}g^{\mp}$ and $\phi
q\overline{q}Q\overline{Q}$ processes have appeared~\cite{Dixon:2009uk}.

In this paper, we calculate the one-loop NMHV amplitude $A^{(1)}_4(\phi,1^+,2^-,3^-,4^-)$ and combine it
with the already known $A^{(1)}_4(\phi^\dagger,1^+,2^-,3^-,4^-)$ amplitude~\cite{Berger:phinite} to
complete the set of analytic helicity amplitudes for the process $0 \rightarrow Hgggg$. The only remaining 
missing analytic piece for the process $pp \rightarrow H + 2 j$ is the $\phi q\overline{q}gg$-NMHV
amplitude.\footnote{We note that the $\phi q\overline{q}gg$-NMHV amplitude has recently been computed~\cite{Badger:2009vh}.}

By employing a four-dimensional unitarity-based strategy, we reconstruct the cut-constructible term  as a
combination of $n$-point scalar integrals (with $n=2,3,4$) which are referred to as   bubble-, triangle-
and box-functions.  By applying a specific set of cuts both to the amplitude and to the basis integrals,
one can isolate specific coefficients appearing in the basis expansion.  Cutting a propagator means
restricting the momenta so that the propagating particle is on-shell, and for four dimensional loop
momenta, setting $n$ propagators on-shell leaves $4-n$ of the loop momentum free.  Hence, in the case of
the quadruple-cuts \cite{Britto:2004nc}, the loop momentum is completely fixed, as result, the
determination of the box-coefficients is reduced to an algebraic operation.  To compute
triangle- and bubble-coefficients, fewer cuts must be applied.  In these cases the free components of the
loop momentum become integration variables of  phase-space integrals which can be carried out with the
mathematical methods of complex analysis. In our calculation we determine the coefficients of the triangle-
and bubble-functions using two variants  of the triple- and double-cut integration techniques
\cite{Forde:intcoeffs,Mastrolia:2006ki,Mastrolia:2009dr}.  

The rational part of the amplitude however, cannot be detected by four-dimensional cuts  and as a result
some other method must be used. We find it convenient to separate the rational part into two terms, one
being sensitive to the number of active light flavours and one which is not. The former piece is
efficiently computed using Feynman diagrams,  whilst the latter can be derived using on-shell recurrence
relations \cite{Britto:rec,Britto:proof}.  

This paper is organised as follows, in section~\ref{sec:effint} we describe the model for Higgs
interactions in the large $m_t$ limit. In particular  we describe the breakdown of Higgs amplitudes into
$\phi$ and $\phi^{\dagger}$ contributions and discuss the colour decomposition of one-loop Higgs plus
multi-gluon amplitudes.   We separate the one-loop primitive amplitude into two parts: $C_4$ which contains
the cut-constructible parts of the amplitude, and $R_4$ which contains the remaining rational pieces.  In
section~\ref{sec:genU} we use various unitarity methods to determine the coefficients of the various
one-loop  basis integrals which appear in the cut-constructible $C_4$ and  we derive the rational term
$R_4$ in section~\ref{sec:rat}.  In section \ref{sec:fullres} we present all the primitive helicity amplitudes
for the $0 \rightarrow Hgggg$ process. Section \ref{sec:num} contains numerical values for the Higgs
helicity amplitudes at a particular phase space point. We draw our conclusions in section \ref{sec:conc}. 
We also enclose two appendices containing the tree expressions used as inputs into the unitarity based
calculations and the definitions of the finite parts of the one-loop basis integrals.  Throughout our
calculations we have made extensive use of the S@M package~\cite{Maitre:2007jq}. 

\section{The Model For Higgs Interactions In The Large $m_t$ Limit}
\label{sec:effint}

We work in an effective theory where the Higgs couples to gluons through a top quark loop, but the top quark degrees of freedom have been integrated out,
\begin{equation}
{\cal L}_H^{\rm int} = \frac{C}{2} H {\rm tr} G_{\mu\nu}G^{\mu\nu}.
\end{equation}
The effective coupling C has been calculated up to order
$\mathcal{O}(\alpha_s^4)$ in
\cite{Chetyrkin:heffalpha3}. However, for our purposes we
need it only up to order $\mathcal{O}(\alpha_s^2)$ \cite{Inami:Heff2l},
\begin{equation}
	C = \frac{\alpha_s}{6\pi v}\left( 1 + \frac{11}{4\pi}\alpha_s \right)+\mathcal{O}(\alpha_s^3)
	\label{eq:effcoupling}
\end{equation}
where $v=246$ GeV and the strong coupling constant is $\alpha_s=g^2/(4\pi)$.

Following Ref.~\cite{Dixon:HiggsMHV} we introduce the complex field $\phi$ (and its conjugate $\phi^\dagger$
\begin{equation}
\phi = \frac{(H+iA)}{2}, \qquad \qquad \phi^\dagger = \frac{(H-iA)}{2}.
\end{equation}
The effective interaction linking gluons and scalar fields also splits into a piece containing  $\phi$ and the self-dual gluon field strengths and another part linking  $\phi^\dagger$ to the anti-self-dual gluon field strengths. The last step conveniently embeds the Higgs interaction
within the MHV structure of the QCD amplitudes. The full Higgs amplitudes are then written as a sum
of the $\phi$ (self-dual) and $\phi^\dagger$ (anti-self-dual) components,
\begin{equation}
	A_n^{(l)}(H;\{p_k\}) = A_n^{(l)}(\phi,\{p_k\})+A_n^{(l)}(\phi^\dagger,\{p_k\}).
\label{eq:H}
\end{equation}
We can also generate pseudo-scalar amplitudes from the difference of $\phi$ and $\phi^{\dagger}$ components,
\begin{equation}
	A_n^{(l)}(A;\{p_k\}) = \frac{1}{i}\left( A_n^{(l)}(\phi,\{p_k\})-A_n^{(l)}(\phi^\dagger,\{p_k\})\right).
\end{equation}
Furthermore parity relates $\phi$ and $\phi^{\dagger}$ amplitudes, 
\begin{eqnarray}
A^{(m)}_n(\phi^{\dagger},g^{\lambda_1}_1,\dots,g^{\lambda_n}_n)=\bigg(A^{(m)}_n(\phi,g^{-\lambda_1}_1,\dots,g^{-\lambda_n}_n)\bigg)^*.
\label{eq:phidagger}
\end{eqnarray}
From now on, we will only consider $\phi$-amplitudes, knowing that all others can be obtained
using eqs.~(\ref{eq:H})--(\ref{eq:phidagger}).

The tree level amplitudes linking a $\phi$ with $n$ gluons 
can be decomposed into colour ordered amplitudes as~\cite{Dawson:Htomultijet,DelDuca:Hto3jets},
\begin{align}
	{\cal A}^{(0)}_n(\phi,\{k_i,\lambda_i,a_i\}) = 
	i C g^{n-2}
	\sum_{\sigma \in S_n/Z_n}
	\tr(T^{a_{\sigma(1)}}\cdots T^{a_{\sigma(n)}})\,
	A^{(0)}_n(\phi,\sigma(1^{\lambda_1},..,n^{\lambda_n})).
	\label{TreeColorDecompositionQ}
\end{align}
Here $S_n/Z_n$ is the group of non-cyclic permutations on $n$
symbols, and $j^{\lambda_j}$ labels the momentum $p_j$ and helicity
$\lambda_j$ of the $j^{\rm th}$ gluon, which carries the adjoint
representation index $a_i$.  The $T^{a_i}$ are fundamental
representation SU$(N_c)$ colour matrices, normalised so that
${\rm Tr}(T^a T^b) = \delta^{ab}$. 
Tree-level amplitudes with a single quark-antiquark pair 
can be decomposed into colour-ordered amplitudes as follows,
\begin{eqnarray}
\lefteqn{
{\cal A}^{(0)}_n(\phi,\{p_i,\lambda_i,a_i\},\{p_j,\lambda_j,i_j\}) }\\
&&= 
i C g^{n-2}
\sum_{\sigma \in S_{n-2}} (T^{a_{\sigma(2)}}\cdots T^{a_{\sigma(n-1)}})_{i_1i_n}\,
A_n(\phi,1^{\lambda},\sigma(2^{\lambda_2},\ldots,{(n-1)}^{\lambda_{n-1}}),
n^{-\lambda})\,,\nonumber 
\label{TreeColorDecomposition}
\end{eqnarray}
where $S_{n-2}$ is the set of permutations of $(n-2)$ gluons.
Quarks are characterised with fundamental colour 
label $i_j$  and
helicity $\lambda_j$ for $j=1,n$.
By current conservation, the quark and antiquark helicities are  related such
that $\lambda_1 = -\lambda_n \equiv \lambda$ where $\lambda = \pm \frac{1}{2}$.

The one-loop amplitudes which are the main subject of this paper
follow the same colour ordering as the pure QCD amplitudes \cite{BDDK:uni1}
and can be decomposed as \cite{Berger:phinite,Badger:1lhiggsallm,Badger:1lphiMHV,Glover:1lphiMHVgeneral},
\begin{align}
	\mc A^{(1)}_n(\phi,\{k_i,\lambda_i,a_i\}) &= i C g^{n}
	\sum_{c=1}^{[n/2]+1}\sum_{\sigma \in S_n/S_{n;c}} G_{n;c}(\sigma)
	A^{(1)}_n(\phi,\sigma(1^{\lambda_1},\ldots,n^{\lambda_n}))
	\label{eq:1lhtogcolour}
\end{align}
where
\begin{align}
	G_{n;1}(1) &= N_c \tr( T^{a_1}\cdots T^{a_n} ) \\
	G_{n;c}(1) &=   \tr( T^{a_1}\cdots T^{a_{c-1} } )
			\tr( T^{a_c}\cdots T^{a_n} )
	\,\, , \, c>2.
	\label{eq:colourfactors}
\end{align}
The sub-leading terms can be computed by summing over various permutations of the leading colour
amplitudes~\cite{BDDK:uni1}. 
 
In this paper we define kinematic invariants associated with sums of gluon momenta as follows, 
\begin{eqnarray}
s_{ij}=(p_i+p_j)^2, \qquad s_{ijk}=(p_i+p_j+p_k)^2 \qquad {\rm etc}.
\end{eqnarray} 
We will express helicity amplitudes using the notation of the spinor-helicity formalism,
\begin{eqnarray}
\la ij \ra &=&\overline{u}_{-}(k_i)u_{+}(k_j),\\
\B ij&=&\overline{u}_{+}(k_i)u_-(k_j),
\end{eqnarray}
where $u_{\pm}(k_i)$ represents a massless Dirac spinor associated with either positive or negative helicity (and a momentum $k_i$). Spinor products are related to kinematic invariants through the following relation,
\begin{eqnarray}
s_{ij}=\la ij \ra[ji]. 
\end{eqnarray}
Chains of spinor products are written as
\begin{eqnarray}
\la i|j|k]=\la ij\ra [jk] \qquad \la i | j k | l \ra = \la ij \ra [jk] \la kl \ra, \qquad  {\rm etc}.
\end{eqnarray}
For example, using momentum conservation we have,
\begin{eqnarray}
 \la i|p_{\phi}|k] = -\sum_{j=1}^{4} \la ij\ra[jk].
 \end{eqnarray}
 
Throughout this paper we use the following expression for  the $\phi$-NMHV tree amplitude
\begin{align}
	&A^{(0)}_n(\phi,1^+,2^-,3^-,4^-) =\nonumber\\ 
&
-\frac{m_\phi^4 \la 24\ra ^4}{s_{124} \la 12\ra  \la 14\ra \la 2|p_\phi|3] \la 4|p_\phi|3]}
+\frac{ \la 4|p_\phi|1]^3}{s_{123} \la 4|p_\phi|3] [12] [23]}
-\frac{ \la 2|p_\phi|1]^3}{s_{134} \la 2|p_\phi|3] [14] [34]}.
\end{align}
This compact form can be derived using the BCFW recursion relations \cite{Britto:rec,Britto:proof} and agrees numerically with the previously
known expression derived from MHV rules \cite{Dixon:HiggsMHV}. It clearly possess the correct symmetry properties under the exchange $\{2\leftrightarrow 4\}$, and factors onto the correct gluon tree amplitude (which is zero) in the limit of vanishing $p_{\phi}$.  The other tree amplitudes we require for this work are listed in Appendix A.

\section{Cut-Constructible Contributions}
\label{sec:genU}

We choose to expand the one-loop primitive amplitude in the following form, 
\begin{eqnarray}
\label{eq:break}
A^{(1)}_4(\phi,1^{+},2^{-},3^{-},4^{-})=c_{\Gamma} (C_4(\phi,1^{+},2^{-},3^{-},4^{-})+R_4(\phi,1^{+},2^{-},3^{-},4^{-})),
\label{eq:bdown}
\end{eqnarray}
where
\begin{equation}
	c_\Gamma = \frac{\Gamma^2(1-\e)\Gamma(1+\e)}{(4\pi)^{2-\e}\Gamma(1-2\e)}.
\end{equation}
In \eqref{eq:break}, $C_4(\phi,1^{+},2^{-},3^{-},4^{-})$ denotes the cut-constructible parts of the amplitude, whilst $R_4(\phi,1^+,2^-,3^-,4^-)$ contains the remaining rational pieces. In this section, we focus our attention on $C_4(\phi,1^{+},2^{-},3^{-},4^{-})$, while an analytic expression for  $R_4(\phi,1^{+},2^{-},3^{-},4^{-})$ is derived in section~\ref{sec:rat}.

We employ the generalised unitarity method
\cite{Britto:2004nc,Forde:intcoeffs,Mastrolia:2006ki,Britto:sqcd,Britto:ccqcd,Mastrolia:2009dr} to calculate the
cut-constructible parts of the one-loop amplitude. We can further decompose $C_4$ in \eqref{eq:bdown}
into a sum over constituent basis integrals,
\begin{eqnarray}
C_4(\phi,1^+,2^-,3^-,4^-)=\sum_{i}{C}_{4;i}\mc{I}_{4;i} + \sum_{i}{C}_{3;i}\mc{I}_{3;i} +\sum_{i}{C}_{2;i}\mc{I}_{2;i}.
\end{eqnarray}
Here $\mc{I}_{j;i}$ represents a $j$-point scalar basis integral, with a coefficient
$C_{j;i}$. The sum over $i$ represents the sum over the partitions of the external
momenta over the $j$ legs of the basis integral. 

Multiple cuts isolate different integral functions and allow the
construction of a linear system of equations from which the coefficients can be extracted. When
considering quadruple cuts of one-loop amplitudes, one is forced to consider complex momenta in
order to fulfill the on-shell constraints~\cite{Britto:2004nc}. The four
on-shell constraints are sufficient to isolate each four-point (box) 
configuration by freezing the loop momentum, thereby allowing the determination of the corresponding coefficient
by a purely algebraic operation.
To isolate the coefficients of lower-point integrals, one needs to cut fewer than four lines.
In this case the loop momenta is no longer completely determined, but, according to the number of cuts, 
some of its components are free variables. 
In this case the computation of the three- (triangle) and two-point (bubble) coefficients can also be reduced 
to algebraic procedures by exploiting the singularity structure of amplitudes in the complex-plane, 
either by explicit subtraction \cite{Ossola:2006us,Berger:2008sj}
or by Laurent expansion \cite{Forde:intcoeffs}.
Alternatively one can extract the coefficients of bubble- and triangle-functions by employing 
the spinor-integration technique, which can be applied to both
double- \cite{Britto:sqcd,Britto:ccqcd} and triple-cuts \cite{Mastrolia:2006ki}.
This method has recently inspired a novel technique for evaluating the double-cut phase-space 
integrals {\it via} Stokes' Theorem applied to functions of 
two complex-conjugated variables \cite{Mastrolia:2009dr}.

\subsection{Box Integral Coefficients}
\label{sec:box}

\begin{figure}[t]
	\begin{center}
		\psfrag{1}{$1$}
		\psfrag{2}{$2$}
		\psfrag{3}{$3$}
		\psfrag{4}{$4$}
		\psfrag{a}{$(a)$}
		\psfrag{b}{$(b)$}
		\psfrag{c}{$(c)$}
		\psfrag{d}{$(d)$}
		\psfrag{p}{$\phi$}
		\psfrag{H4}{\hspace{-3mm}$C_{4;\phi 4|1|2|3}$}
		\psfrag{Hx1}{\hspace{-5mm}$C_{4;\phi|1|2|34}$}
		\psfrag{2xH}{\hspace{-3mm}$C_{4;\phi|34|1|2}$}
		\psfrag{4xHx1}{\hspace{-3mm}$C_{4;\phi|1|23|4}$}
		\includegraphics[width=\textwidth]{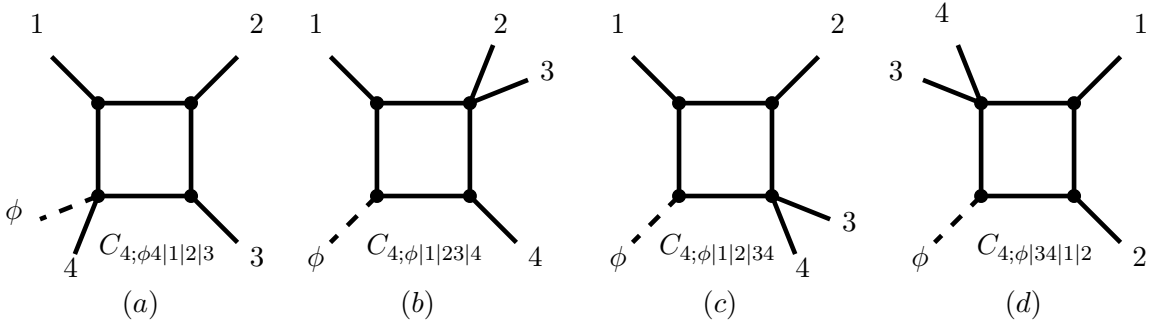}
	\end{center}
	\caption{The various box integral topologies that appear for $A_4^{(1)}(\phi,1,2,3,4)$. From the four topologies we
	must also include cyclic permutations of the four gluons. Here $(a)$ has one off-shell leg (one-mass) whilst $(b)$-$(d)$ have two off-shell
	legs. In $(b)$ the two off-shell legs are not adjacent and we refer to this configuration to as the two-mass easy box,  while in $(c)$ and $(d)$ the two off-shell legs are adjacent and we label them as   two-mass hard boxes.}
	\label{fig:boxbasis}
\end{figure}

We begin our calculation of the $\phi$-NMHV amplitude by computing the coefficients of the scalar
boxes using
generalised unitarity with complex momenta  \cite{Britto:2004nc}.  In general there are sixteen box
topologies, which can be obtained from cyclic permutations of those shown in Fig.~\ref{fig:boxbasis}.  
When a box graph contains a sequence of three point vertices, a non-vanishing solution is only found when the vertices alternate between MHV and $\overline{\rm{MHV}}$-types~\cite{Britto:2004nc}.
For
the specific helicity configuration we consider this is not possible for the graphs of Fig~\ref{fig:boxbasis}$(b)$; the three-point MHV vertex involving $\phi$ and two loop-gluons cannot have both adjacent three-point gluon vertices to be of $\overline{\rm{MHV}}$-type.  Therefore,  the coefficients of the two-mass easy boxes (Fig~\ref{fig:boxbasis}$(b)$) are all
zero.  

Of the remaining 12 coefficients, a further 5 are related to each other by the $\{2\leftrightarrow 4\}$ symmetry,
\begin{align}
\label{eq:boxcoeffs0}
{C}_{4;\phi4|1|2|3}(\phi,1^+,2^-,3^-,4^-)&={C}_{4;\phi2|3|4|1}(\phi,1^+,4^-,3^-,2^-),\\
C_{4;\phi|23|4|1}(\phi,1^+,2^-,3^-,4^-)&={C}_{4;\phi|1|2|34}(\phi,1^+,4^-,3^-,2^-),\\
C_{4;\phi|34|1|2}(\phi,1^+,2^-,3^-,4^-)&={C}_{4;\phi|4|1|23}(\phi,1^+,4^-,3^-,2^-),\\
C_{4;\phi|12|3|4}(\phi,1^+,2^-,3^-,4^-)&={C}_{4;\phi|2|3|41}(\phi,1^+,4^-,3^-,2^-),\\
{C}_{4;\phi|3|4|12}(\phi,1^+,2^-,3^-,4^-)&={C}_{4;\phi|41|2|3}(\phi,1^+,4^-,3^-,2^-).
\end{align}
We find that two of the one-mass box coefficients (Fig~\ref{fig:boxbasis}$(a)$) are given by, 
\begin{eqnarray}
\label{eq:boxcoeffs1}
C_{4;\phi1|2|3|4}(\phi,1^+,2^-,3^-,4^-)&=&
\frac{s_{23}s_{34}s_{234}^3}{2\la1|p_{\phi}|2]\la1|p_{\phi}|4][23][34]},\\
\label{eq:boxcoeffs}
C_{4;\phi2|3|4|1}(\phi,1^+,2^-,3^-,4^-)&=&
\frac{s_{34}s_{41}\la2|p_{\phi}|1]^3}{2s_{134}\la2|p_{\phi}|3][34][41]}
+\frac{s_{34}s_{41}\la34\ra^3m_{\phi}^4}{2s_{134}\la1|p_{\phi}|2]\la3|p_{\phi}|2]\la41\ra}.
\end{eqnarray}
We also find that three of the two-mass hard boxes (Fig.~\ref{fig:boxbasis}$(d)$) have coefficients related to the coefficients of eqs.~\eqref{eq:boxcoeffs0}, \eqref{eq:boxcoeffs1} and \eqref{eq:boxcoeffs},
\begin{align}
{C}_{4;\phi|12|3|4}(\phi,1^+,2^-,3^-,4^-)&=\frac{s_{123}s_{34}}{s_{23}s_{12}}{C}_{4;\phi4|1|2|3}(\phi,1^+,2^-,3^-,4^-),\\
{C}_{4;\phi|23|4|1}(\phi,1^+,2^-,3^-,4^-)&=\frac{s_{234}s_{41}}{s_{23}s_{34}}{C}_{4;\phi1|2|3|4}(\phi,1^+,2^-,3^-,4^-),\\
{C}_{4;\phi|34|1|2}(\phi,1^+,2^-,3^-,4^-)&=\frac{s_{134}s_{12}}{s_{41}s_{34}}{C}_{4;\phi2|3|4|1}(\phi,1^+,2^-,3^-,4^-).
\label{eq:boxcoeffs2}
\end{align}
The final two-mass hard box coefficient is, \newpage
\begin{eqnarray}
\label{eq:boxcoeffs3}
{C}_{4;\phi|3|4|12}(\phi,1^+,2^-,3^-,4^-)&=&\frac{s_{34}}{2}\bigg(\frac{\la3|p_{\phi}|1]^4}{\la3|p_{\phi}|2]\la3|p_{\phi}|4][21][41]}+\frac{\la24\ra^4m_{\phi}^4}{\la12\ra\la14\ra\la2|p_{\phi}|3]\la4|p_{\phi}|3]}\bigg)\nonumber\\
\end{eqnarray}
The remaining  one-mass box configuration $C_{4;\phi 3|4|1|2}$ is the only one which receives contributions from $N_f$ fermionic and $N_s$ scalar
loops,
\begin{eqnarray}
	\lefteqn{{C}_{4;\phi3|4|1|2}(\phi,1^+,2^-,3^-,4^-)=
	s_{41}s_{12}\bigg(\frac{1}{s_{124}s_{34}}{C}_{4;\phi|3|4|12}(\phi,1^+,2^-,3^-,4^-)}\nonumber\\
	&&-\left(1-\frac{N_f}{4N_c}\right)\frac{2\AB3{p_\phi}1^2}{s_{124}\B24^2}
	-\left(1-\frac{N_f}{N_c}+\frac{N_s}{N_c}\right)
	\frac{\B12\B41\AB3{p_\phi}2\AB3{p_\phi}4}{s_{124}\B24^4}\bigg).
	\label{eq:C4H3full}
\end{eqnarray}
Each of the coefficients \eqref{eq:boxcoeffs}, \eqref{eq:boxcoeffs1}, \eqref{eq:boxcoeffs3} and \eqref{eq:C4H3full} correctly tends to zero in the soft Higgs limit ($p_{\phi}\rightarrow 0)$. 

\subsection{Triangle Integral Coefficients}

\begin{figure}[h]
	\begin{center}
		\psfrag{Hx12x34}{$C_{3;\phi|12|34}$}
		\psfrag{Hx1x234}{$C_{3;\phi|1|234}$}
		\psfrag{Hx123x4}{$C_{3;\phi|123|4}$}
		\psfrag{H34x1x2}{$C_{3;\phi 34|1|2}$}
		\psfrag{H4x12x3}{$C_{3;\phi 4|12|3}$}
		\psfrag{H4x1x23}{$C_{3;\phi 4|1|23}$}
		\psfrag{a}{$(a)$}
		\psfrag{b}{$(b)$}
		\psfrag{c}{$(c)$}
		\psfrag{d}{$(d)$}
		\psfrag{e}{$(e)$}
		\psfrag{f}{$(f)$}
		\psfrag{1}{$1$}
                \psfrag{2}{$2$}
                \psfrag{3}{$3$}
                \psfrag{4}{$4$}
                \psfrag{P}{$\phi$}
                \includegraphics[width=0.7\textwidth]{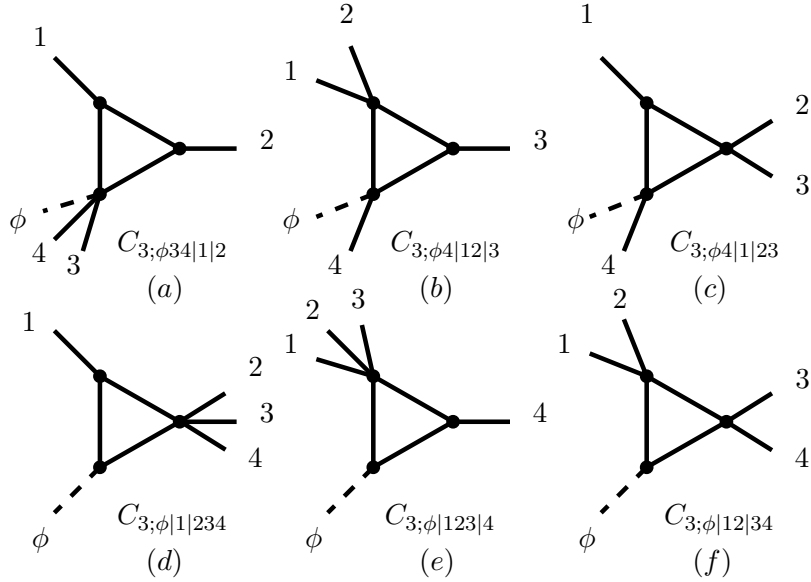}
	\end{center}
	\caption{The various triangle integral topologies that appear for $A_4^{(1)}(\phi,1,2,3,4)$. From the six topologies we
	must also include cyclic permutations of the four gluons.   $(a)$
has one off-shell leg, $(b)$-$(e)$ have two off-shell legs while in 
   $(f)$ all legs are off-shell.}
	\label{fig:tribasis}
\end{figure}
Altogether, there are twenty-four triangle
topologies, which can be obtained from cyclic permutations of those shown in Fig.~\ref{fig:tribasis}.
The different topologies can be characterised by the number of off-shell legs.   Fig.~\ref{fig:tribasis}$(a)$
has one off-shell leg, Figs.~\ref{fig:tribasis}$(b)$-$(e)$ have two off-shell legs while for 
   Fig.~\ref{fig:tribasis}$(f)$ all legs are off-shell.
We refer to the triangle integrals with one- and two-off-shell legs as one-mass and two-mass respectively. They have the following form,
\begin{eqnarray}
{\cal I}^{1m}_3(s) \propto \frac{1}{\e^2}\frac{1}{s}\bigg(\frac{\mu^2}{-s}\bigg)^{\e}, \qquad \qquad  {\cal I}^{2m}_3(s,t) \propto \frac{1}{\e^2}\frac{1}{(s-t)}\bigg(\bigg(\frac{\mu^2}{-s}\bigg)^{\e}-\bigg(\frac{\mu^2}{-t}\bigg)^{\e}\bigg) 
\end{eqnarray}
and therefore only contribute pole pieces in $\e$ to the overall amplitude. In fact, the sole role of these functions is to ensure the correct infrared behaviour by combining with the box pieces to generate the following pole structure, 
\begin{align}
	A^{(1)}(\phi,1^+,2^-,3^-,4^-)  = -
	A^{(0)}(\phi,1^+,2^-,3^-,4^-)\frac{c_\Gamma}{\e^2}&
	\sum_{i=1}^4 \left( \frac{\mu^2}{-s_{i i+1}}\right)^\e + {\cal O}(\e^0).
	\label{eq:IR}
\end{align}
This relation holds for arbitrary external helicities \cite{Badger:1lphiMHV,Badger:1lhiggsallm,Glover:1lphiMHVgeneral}. We computed the coefficients of all one- and two-mass triangles and explicitly verified eq.~\eqref{eq:IR}. The non-trivial relationship between the triangle and box coefficients provides a strong check of our calculation. However, we find it more compact to present the final answer in a basis free of one- and two-mass triangles. That is, we choose to expand the box integral functions into divergent and finite pieces, combining the divergent pieces with the one- and two- mass triangles to form~\eqref{eq:IR} and giving explicit results for the finite pieces of the box functions. 

A new feature in the $\phi$-NMHV amplitudes is the presence of  three-mass triangles, shown in Fig.~\ref{fig:tribasis}$(f)$. In previous calculations~\cite{Berger:phinite,Badger:1lphiMHV,Badger:1lhiggsallm,Glover:1lphiMHVgeneral,Dixon:2009uk} the external gluon helicities prevented these contributions
from occurring.
 
There are four three-mass triangles, which satisfy,
\begin{align}
C_{3;\phi|34|12}(\phi,1^+,2^-,3^-,4^-) &= C_{3;\phi|12|34}(\phi,1^+,2^-,3^-,4^-)\\
C_{3;\phi|41|23}(\phi,1^+,2^-,3^-,4^-) &= C_{3;\phi|23|41}(\phi,1^+,2^-,3^-,4^-).
\end{align}
The symmetry under the exchange of gluons with momenta $p_2$ and $p_4$ relates the remaining two coefficients, 
\begin{align}
		C_{3;\phi|23|41}(\phi,1^+,2^-,3^-,4^-) &=C_{3;\phi|12|34}(\phi,1^+,4^-,3^-,2^-). 
\end{align}		
To compute $C_{3;\phi|23|41}$ we use both Forde's method \cite{Forde:intcoeffs} 
and the spinor integration technique \cite{Mastrolia:2006ki}. 
For a given triangle coefficient $C_{3;K_1|K_2|K_3}(\phi,1^+,2^-,3^-,4^-)$ with off-shell momenta $K_1$, $K_2$ and $K_3$,
we introduce the following massless projection vectors
\begin{align}
	\kfm1 &= \gamma\frac{\gamma K_1^{\mu}-K_1^2K_2^{\mu}}{\gamma^2-K_1^2K_2^2}, \nonumber \\
	\kfm2 &= \gamma\frac{\gamma K_2^{\mu}-K_2^2K_1^{\mu}}{\gamma^2-K_1^2K_2^2}, \nonumber \\
	\gamma_\pm(K_1,K_2) &= K_1\cdot K_2 \pm \sqrt{ K_1\cdot K_2^2-K_1^2K_2^2}.
\label{eq:flatdef}
\end{align}
In terms of these quantities we find,
\begin{align}
	C_{3;\phi|12|34}(\phi,1^+,2^-,3^-,4^-) &= \sum_{\gamma=\gamma_\pm(p_\phi,p_1+p_2)}
	-\frac{m_\phi^4\A{\kf1}{2}^3\A 34^3}{2\gamma(\gamma+m_\phi^2)\A{\kf1}{1}\A{\kf1}{3}\A{\kf1}{4}\A 12},
	\end{align}
which, as expected, correctly vanishes in the soft Higgs limit ($p_{\phi}\rightarrow 0)$. 

\subsection{Bubble Integral Coefficients}
\label{sec:bub}

\begin{figure}[h]
	\begin{center}
        \psfrag{H}{$C_{2;\phi}$}
        \psfrag{H4}{\hspace{-2mm}$C_{2;\phi 4}$}
        \psfrag{H34}{\hspace{-2mm}$C_{2;\phi 34}$}
		\psfrag{a}{$(a)$}
		\psfrag{b}{$(b)$}
		\psfrag{c}{$(c)$}
		\psfrag{1}{$1$}
                \psfrag{2}{$2$}
                \psfrag{3}{$3$}
                \psfrag{4}{$4$}
                \psfrag{P}{$\phi$}
		\includegraphics[width=0.9\textwidth]{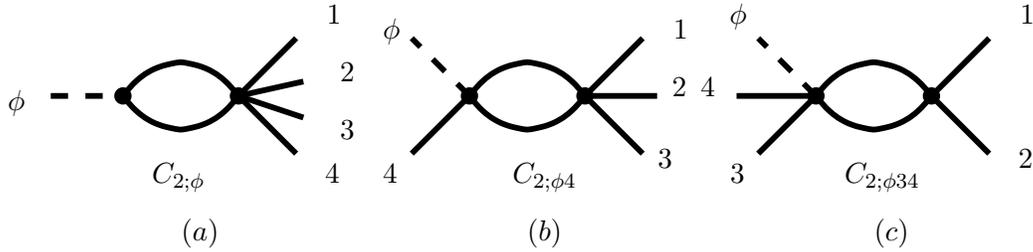}
	\end{center}
	\caption{The three bubble integral topologies that appear for $A_4^{(1)}(\phi,1,2,3,4)$. We
	must also include cyclic permutations of the four gluons.}
	\label{fig:bubs}
\end{figure}
The non-vanishing bubble topologies for the $\phi$-NMHV amplitude are shown in Fig.~\ref{fig:bubs}. We find that the double-cuts associated with Fig.~\ref{fig:bubs}$(a)$ contain only contributions from boxes and triangles, and therefore the coefficient of $\log(s_{1234})$ is zero. In a similar fashion, the double cuts associated with diagram Fig.~\ref{fig:bubs}$(c)$ with two external gluons with negative helicity emitted from  the right hand vertex have only box and triangle contributions, so that the coefficients of $\log({s_{23}})$ and $\log({s_{34}})$ are also zero.  

The leading singularity of the bubble integral is ${\cal O}(1/\e)$,
\begin{eqnarray}
{\cal I}_2(s) \propto \frac{1}{(1-2\e)\e}\bigg(\frac{\mu^2}{-s}\bigg)^{\e}.
\end{eqnarray} 
However for the total amplitude there is no overall $\e$ pole, and this implies a relation amongst the bubble coefficients such that,
\begin{equation}
	\sum_{k=1}^4 \left ( C_{2;\phi k}+C_{2;\phi k k+1} \right )= 0.	
\end{equation}
It is therefore most natural to work with $\log$'s of ratios of kinematic scales (rather than $\log(s/\mu^2)$), since the coefficients of individual logarithms must cancel pairwise. To this end, we express our result in terms of the following functions, 
\begin{eqnarray}
L_k(s,t)=\frac{\log{(s/t)}}{(s-t)^k}.
\end{eqnarray}

Using the Stokes' theorem method \cite{Mastrolia:2009dr}, we generated compact analytic expressions for the coefficients of each bubble-function, which we also checked numerically with Forde's method \cite{Forde:intcoeffs}. 
The combination of all double-cuts is given by,
\begin{eqnarray}
C_{2}(\phi,1^+,2^-,3^-,4^-) = \left(4-\frac{N_f}{N}\right)C^{(1)}_{2}+\left(1-\frac{N_f}{N_c}+\frac{N_s}{N_c}\right)C^{(2)}_{2}
\end{eqnarray}
with
\begin{align}
C^{(1)}_{2} = 
-\left\{
\frac{\la24\ra\la3|p_\phi|1]^2}{s_{124}[42]}L_1\left(s_{124},s_{12}\right)
-\frac{\la23\ra\la4|p_\phi|1]^2}{s_{123}[32]}L_1\left(s_{123},s_{12}\right)
\right\}- \biggl \{ (2 \leftrightarrow 4 )\biggr \}
\end{align}
and
\begin{align}
&C^{(2)}_{2} = -\bigg\{\frac{2s_{124}\la24\ra\la34\ra^2[41]^2}{3[42]}L_3\left(s_{124},s_{12}\right)\nonumber\\&+\frac{\la34\ra[41]\left(3s_{124}\la34\ra[41]+\la24\ra\la3|p_\phi|1][42]\right)}{3[42]^2}L_2\left(s_{124},s_{12}\right)\nonumber\\&
+\left(\frac{2s_{124}\la34\ra^2[41]^2}{\la24\ra[42]^3}-\frac{\la24\ra\la3|p_\phi|1]^2}{3s_{124}[42]}\right)L_1\left(s_{124},s_{12}\right)\nonumber\\&
+\frac{\la3|p_\phi|1]\left(4s_{124}\la34\ra[41]+\la3|p_\phi|1](2s_{14}+s_{24})\right)}{s_{124}\la24\ra[42]^3}L_0\left(s_{124},s_{12}\right)\nonumber\\&
-\frac{2s_{123}\la23\ra\la34\ra^2[31]^2}{3[32]}L_3\left(s_{123},s_{12}\right)
+\frac{\la23\ra\la34\ra[31]\la4|p_\phi|1]}{3[32]}L_2\left(s_{123},s_{12}\right)\nonumber\\&
+\frac{\la23\ra\la4|p_\phi|1]^2}{3s_{123}[32]}L_1\left(s_{123},s_{12}\right)
\bigg\}- \bigg\{(2 \leftrightarrow 4 )\bigg\}.
\label{eq:bub2}
\end{align}
In the above formulae (and those following) we stress that the symmetrising action applies to the entire formula, and also acts on the kinematic invariants of the basis functions.  
We see that $C_2(\phi,1^+,2^-,3^-,4^-)$ vanishes in  the soft Higgs limit $p_{\phi}\rightarrow 0$. 

\subsection{The Cut-Completion terms}
\label{sec:CRrat}

The basis functions $L_3(s,t)$ and $L_2(s,t)$ are singular as $s \rightarrow t$. Since this is an unphysical limit one expects to find some cut-predictable rational pieces which 
ensure the correct behaviour of the amplitude as these quantities approach each other. These  rational pieces are called the cut-completion terms and are obtained by making the following replacements in~\eqref{eq:bub2}
\begin{eqnarray}
L_3(s,t) \to \hat{L}_3(s,t)&=&L_3(s,t)-\frac{1}{2(s-t)^2}\bigg(\frac{1}{s}+\frac{1}{t}\bigg), \nonumber \\
L_2(s,t) \to \hat{L}_2(s,t)&=&L_2(s,t)-\frac{1}{2(s-t)}\bigg(\frac{1}{s}+\frac{1}{t}\bigg), \nonumber \\
L_1(s,t) \to \hat{L}_1(s,t)&=&L_1(s,t),\nonumber \\
L_0(s,t) \to \hat{L}_0(s,t)&=&L_0(s,t).
\label{eq:lhat}
\end{eqnarray}

\section{Rational Terms}
\label{sec:rat}

We now turn our attention to the calculation of the remaining rational
part of the amplitude.  In general the cut-unpredictable rational
part of $\phi$ plus gluon amplitudes contains two types of pieces, 
a homogeneous piece, which is
insensitive to the number of active flavours and a piece proportional to $(1-N_f/N_c+N_s/N_c)$,
\begin{eqnarray}
R_4(\phi,1^+,2^-,3^-,4^-)=R^h_4(\phi,1^+,2^-,3^-,4^-)+\bigg(1-\frac{N_f}{N_c}+\frac{N_s}{N_c}\bigg)R^{N_P}_4(\phi,1^+,2^-,3^-,4^-).\nonumber\\
\end{eqnarray}
The homogeneous term $R^h_4(\phi,1^+,2^-,3^-,4^-)$ can be simply calculated using the BCFW recursion relations \cite{Britto:rec,Britto:proof},
\begin{eqnarray}
R^{h}_4(\phi,1^+,2^-,3^-,4^-)=2A^{(0)}(\phi,1^+,2^-,3^-,4^-). 
\end{eqnarray}
This contribution cancels against a similar homogeneous term for the $\phi^{\dagger}$ amplitude when combining the $\phi$ and $\phi^\dagger$ amplitudes to form the Higgs amplitude.

The $N_P$ piece allows the propagation of quarks (or scalars) in the loop, and can be completely reconstructed by considering only the fermion (scalar) loop contribution.  Furthermore, one can extract the $\phi$ contribution to $R_4^{N_P}$ 
by considering the full Higgs amplitude and removing the fully rational $\phi^\dagger$ contribution calculated in~\cite{Berger:phinite}.
Since there is no direct $Hq \overline{q}$ coupling in the effective theory, 
the most complicated structure is a second-rank tensor box
configuration.  Of the 739 diagrams contributing to the $Hgggg$ amplitude\footnote{Feynman diagrams were generated with the aid of {\tt QGRAF} \cite{Nogueira:qraf}.}, only 136 contain fermion loops and are straightforward to evaluate.

After subtracting the cut-completion and homogeneous rational terms from the explicit Feynman diagram calculation the following rational pieces remain. 
\begin{eqnarray}
R^{N_P}_4(H,1^+,2^-,3^-,4^-)=&&\bigg\{\frac{1}{2}\bigg(\frac{\la 23\ra\la34\ra\la4|p_{H}|1][31]}{3s_{123}\la12\ra[21][32]}
-\frac{\la3|p_{H}|1]^2}{s_{124}[42]^2}+\frac{\la24\ra\la34\ra\la3|p_{H}|1][41]}{3s_{124}s_{12}[42]}\nonumber\\&&
-\frac{[12]^2\la23\ra^2}{s_{14}[42]^2}-\frac{\la24\ra(s_{23}s_{24}+s_{23}s_{34}+s_{24}s_{34})}{3\la12\ra\la14\ra[23][34][42]}\nonumber\\&&
+\frac{\la2|p_{H}|1]\la4|p_{H}|1]}{3s_{234}[23][34]}-\frac{2[12]\la23\ra[31]^2}{3[23]^2[41][34]}\bigg)\bigg\}  + \bigg\{(2\leftrightarrow 4) \bigg\} .
\label{eq:Hrat}
\end{eqnarray}
The last line in the above equation is the one-loop rational expression for the $\phi^{\dagger}$ contribution~\cite{Berger:phinite}. We can thus define the rational terms for the 
$\phi$ contribution. 
\begin{eqnarray}
R^{N_P}_4(\phi,1^+,2^-,3^-,4^-)=&&\bigg\{\frac{1}{2}\bigg(\frac{\la 23\ra\la34\ra\la4|p_{H}|1][31]}{3s_{123}\la12\ra[21][32]}
-\frac{\la3|p_{H}|1]^2}{s_{124}[42]^2}+\frac{\la24\ra\la34\ra\la3|p_{H}|1][41]}{3s_{124}s_{12}[42]}\nonumber\\&&
-\frac{[12]^2\la23\ra^2}{s_{14}[42]^2}-\frac{\la24\ra(s_{23}s_{24}+s_{23}s_{34}+s_{24}s_{34})}{3\la12\ra\la14\ra[23][34][42]}\bigg)\bigg\}  + \bigg\{(2\leftrightarrow 4) \bigg\}. \nonumber \\
\label{eq:phirat}
\end{eqnarray}

\section{Higgs plus four gluon amplitudes }
\label{sec:fullres}

In this section we present complete expressions for the one-loop amplitudes needed to calculate the process $0 \rightarrow Hgggg$ at NLO. 

The one-loop amplitudes presented here are computed in the four-dimensional
helicity scheme and are not renormalised.   To perform an $\overline{MS}$
renormalisation, one should subtract an $\overline{MS}$ counterterm from
$A^{(1)}_4$,
\begin{equation}
A^{(1)}_4 \to A^{(1)}_4 - c_\Gamma 2\frac{\beta_0}{\epsilon}
A^{(0)}_4.
\end{equation}
The Wilson coefficient \eqref{eq:effcoupling} produces an additional finite contribution,
\begin{equation}
A^{(1)}_4 \to A^{(1)}_4 +c_\Gamma \,11\,A^{(0)}_4.
\end{equation}

We choose to split the un-renormalised amplitude into (completed) cut-constructible pieces and rational terms. We also separate the infra-red divergent and finite parts of the amplitude. The basis functions for the finite part of the cut-constructable pieces are  one-mass and two-mass boxes, three-mass triangles, and completed functions $\hat{L}_i(s,t)$ of eq.~\eqref{eq:lhat}.
We define the finite pieces of the box and three-mass triangle integrals in Appendix B.  

We express a generic helicity configuration in the following form 
\begin{eqnarray}
A^{(1)}_4(H,1^{\lambda_1},2^{\lambda_2},3^{\lambda_3},4^{\lambda_4})=c_{\Gamma} (C_4(H,1^{\lambda_1},2^{\lambda_2},3^{\lambda_3},4^{\lambda_4})+R_4(H,1^{\lambda_1},2^{\lambda_2},3^{\lambda_3},4^{\lambda_4})),\nonumber\\
\end{eqnarray}
where $C_4$ represents the cut-constructible part of the amplitude and $R_4$ the rational pieces. 
We further separate $C_4(H,1^{\lambda_1},2^{\lambda_2},3^{\lambda_3},4^{\lambda_4})$ into divergent and finite pieces,
\begin{eqnarray}
C_4(H,1^{\lambda_1},2^{\lambda_2},3^{\lambda_3},4^{\lambda_4})=V_4(H,1^{\lambda_1},2^{\lambda_2},3^{\lambda_3},4^{\lambda_4})+F_4(H,1^{\lambda_1},2^{\lambda_2},3^{\lambda_3},4^{\lambda_4}).
\end{eqnarray}
The divergent part $V_4$ contain the $\epsilon$ singularities generated by the box and triangle contributions, and which satisfy the helicity independent infrared singularity condition,
\begin{eqnarray}
V_4(H,1^{\lambda_1},2^{\lambda_2},3^{\lambda_3},4^{\lambda_4})=-A^{(0)}(H,1^{\lambda_1},2^{\lambda_2},3^{\lambda_3},4^{\lambda_4})\frac{1}{\epsilon^2}\left(\sum_{i=1}^{4}\left(\frac{\mu^2}{-s_{i(i+1)}}\right)^\e\right).
\end{eqnarray}
The remaining cut-constructible and rational terms are finite, and depend non-trivially on the helicity configuration of the gluons.

\subsection{The all-minus amplitude $A^{(1)}_4(H,1^-,2^-,3^-,4^-)$}
The all-minus amplitude is symmetric under cyclic permutations of the four gluons.
The finite part (of the cut-constructible piece) is~\cite{Badger:1lhiggsallm}, 
\begin{eqnarray}
F_4(H,1^-,2^-,3^-,4^-)&&=
\bigg\{-\frac{m_H^4}{2[12][23][34][41]}\bigg(\frac{1}{2}\FFtme(s_{123},s_{234};m_H^2,s_{23})\nonumber\\&&+\frac{1}{2}\FFtme(s_{123},s_{124};m_H^2,s_{12})+\FFom(s_{23},s_{34};s_{234})\bigg)\bigg\}\nonumber\\&&
+\bigg\{ (1\leftrightarrow 4), (2\leftrightarrow 3) \bigg\}+\bigg\{ (1\leftrightarrow 2), (3\leftrightarrow 4) \bigg\}+\bigg\{ (1\leftrightarrow 3), (2\leftrightarrow 4) \bigg\}\nonumber\\&&
\end{eqnarray}
while the rational part is given by~\cite{Badger:1lhiggsallm,Berger:phinite}
\begin{eqnarray}
R_4(H,1^-,2^-,3^-,4^-)&=&\bigg\{\frac{1}{3}\bigg(1-\frac{N_f}{N_c}+\frac{N_s}{N_c}\bigg)\bigg(-\frac{s_{13}\la4|P_{H}|2]^2}{s_{123}[12]^2[23]^2}
+\frac{\la34\ra^2}{[12]^2}+2\frac{\la34\ra\la41\ra}{[12][23]}\nonumber\\&&+\frac{s_{12}s_{34}+s_{123}s_{234}-s_{12}^2}{2[12][23][34][41]}\bigg)\bigg\}
+{\rm ~cyclic~permutations}.
\end{eqnarray}

\subsection{The MHV amplitude $A^{(1)}_4(H,1^-,2^-,3^+,4^+)$}
For the MHV amplitude with adjacent negative helicity gluons  there is an overall ($(1 \leftrightarrow 2)$,$(3 \rightarrow 4)$) symmetry. 
The finite cut-constructible part is~\cite{Badger:1lphiMHV},
\begin{eqnarray}
F_4(H,1^-,2^-,3^+,4^+)&=&\bigg\{\bigg[-\frac{\la12\ra^3}{2\la23\ra\la34\ra\la41\ra}\bigg(\FFtme(s_{123},s_{234};m_H^2,s_{23})\nonumber\\&&
+\frac{1}{2}\FFtme(s_{234},s_{134};m_H^2,s_{34})+\frac{1}{2}\FFtme(s_{124},s_{123};m_H^2,s_{12})\nonumber\\&&+\FFom(s_{23},s_{34};s_{234})
+\FFom(s_{14},s_{12};s_{124})\bigg)\nonumber\\&&
-4\bigg(1-\frac{N_f}{4N_c}\bigg)
\frac{\la12\ra^2[43]}{\la34\ra}\hat{L}_1(s_{134},s_{14})\nonumber\\&&
-\bigg(1-\frac{N_f}{N_c} +\frac{N_s}{N_c}\bigg)\bigg(\frac{[43]\la13p_{H}2\ra(\la13p_{H}2\ra+\la1432\ra)}{3\la34\ra}\hat{L}_3(s_{134},s_{14})\nonumber\\&&
-\frac{\la12\ra^2[43]}{3\la34\ra}\hat{L}_1(s_{134},s_{14})\bigg)\bigg]+\bigg[(1\leftrightarrow 3),(2\leftrightarrow4)\bigg]_{\la ij \ra \leftrightarrow [ij]} \bigg\}\nonumber\\&&+\bigg\{(1\leftrightarrow 2),(3\leftrightarrow4) \bigg\}.
\end{eqnarray}
The rational terms $R_4$ have the same symmetries~\cite{Badger:1lphiMHV}, 
\begin{eqnarray}
R_4(H,1^-,2^-,3^+,4^+)&=&\bigg\{\bigg[\bigg(1-\frac{N_f}{N_c}+\frac{N_s}{N_c}\bigg)\frac{[34]}{3\la34\ra}\bigg(
-\frac{\la23\ra \la1|p_{H}|3]^2}{\la34\ra[43][32]s_{234}}
-\frac{\la14\ra\la3|P_{12}|3]}{\la34\ra[12][32]}\nonumber\\&&+\frac{\la12\ra^2}{2\la34\ra[43]}-\frac{\la12\ra}{2[12]}-\frac{\la12\ra\la2|P_{13}|4]}{2[41]s_{341}}+\frac{\la12\ra^2}{2s_{41}}\bigg)\bigg]\nonumber\\&&+\bigg[(1\leftrightarrow 3),(2\leftrightarrow4)\bigg]_{\la ij \ra \leftrightarrow [ij]} \bigg\}+\bigg\{(1\leftrightarrow 2),(3\leftrightarrow4) \bigg\}.
\end{eqnarray}

\subsection{The MHV amplitude $A^{(1)}_4(H,1^-,2^+,3^-,4^+)$}
The alternating helicity MHV configuration has the larger set of symmetries, $(1 \leftrightarrow 3)$, $(2\leftrightarrow 4)$ and $((1\leftrightarrow 3),(2\leftrightarrow 4))$.  The finite cut-constructible contribution is~\cite{Glover:1lphiMHVgeneral},
\begin{eqnarray}
F_4(H,1^-,2^+,3^-,4^+)&=&\bigg\{\bigg[-\frac{\la13\ra^4}{2\la12\ra\la23\ra\la34\ra\la41\ra}\bigg(\FFtme(s_{123},s_{234};m_H^2,s_{23})\nonumber\\&&+\frac{1}{2}\FFom(s_{23},s_{34};s_{234})+\frac{1}{2}\FFom(s_{34},s_{14};s_{134})\bigg)\nonumber\\&&
+4\bigg(1-\frac{N_f}{4N_c}\bigg)\bigg(-\frac{\la13\ra^2}{\la24\ra}\bigg(\frac{1}{4\la24\ra}\FFom(s_{23},s_{34};s_{234})\nonumber\\&&-[42]\hat{L}_1(s_{234},s_{23})\bigg)\bigg)
+2\bigg(1-\frac{N_f}{N_c}+\frac{N_s}{N_c}\bigg)\bigg(-\frac{\la12\ra\la41\ra\la23\ra\la34\ra}{\la24\ra^3}\nonumber\\&&\times
\bigg(\frac{1}{4\la24\ra}\FFom(s_{23},s_{34};s_{234})-[42]\hat{L}_1(s_{234},s_{23})\bigg)\nonumber\\&&
-\frac{\la23\ra\la41\ra[42]^2}{\la24\ra}\bigg(\frac{\la14\ra\la23\ra[42]}{3}\hat{L}_3(s_{234},s_{23})\nonumber\\&&-\frac{\la12\ra\la34\ra}{2\la24\ra}\hat{L}_2(s_{234},s_{23})\bigg)\bigg)\bigg]
+\bigg[ (1\leftrightarrow 2),(3 \leftrightarrow 4)\bigg]_{\la ij \ra \leftrightarrow [ij]}\bigg\}\nonumber\\&&
+\bigg\{ (1\leftrightarrow 3)\bigg\}+\bigg\{ (2\leftrightarrow 4)\bigg\}+\bigg\{ (1\leftrightarrow 3), (2\leftrightarrow 4)\bigg\}
\end{eqnarray}
while the rational part is given by~\cite{Glover:1lphiMHVgeneral},
\begin{eqnarray}
R_4(H,1^-,2^+,3^-,4^+)&=&\bigg\{\bigg[-\bigg(1-\frac{N_f}{N_c}+\frac{N_s}{N_c}\bigg)\frac{[24]^4}{12[12][23][34][41]}
\bigg(\frac{s_{23}s_{34}}{s_{24}s_{124}}-3\frac{s_{23}s_{34}}{s_{24}^2}\bigg)\bigg]\nonumber\\&&+\bigg[ (1\leftrightarrow 2),(3 \leftrightarrow 4)\bigg]_{\la ij \ra \leftrightarrow [ij]}\bigg\}\nonumber\\&&
+\bigg\{ (1\leftrightarrow 3)\bigg\}+\bigg\{ (2\leftrightarrow 4)\bigg\}+\bigg\{ (1\leftrightarrow 3), (2\leftrightarrow 4)\bigg\}.
\label{eq:phiTotRat}
\end{eqnarray}

\subsection{The NMHV amplitude $A^{(1)}_4(H,1^+,2^-,3^-,4^-)$}
By combining the results for the NMHV $\phi$ amplitudes given in sections~\ref{sec:genU} and \ref{sec:rat} and the rational $\phi^\dagger$ amplitude of ~\cite{Berger:phinite} according to eq.~\eqref{eq:H}, we obtain the Higgs NMHV-amplitude, which
is symmetric under the exchange $(2 \leftrightarrow 4)$.
The finite cut-constructible contribution is,
\begin{eqnarray}
F_4(H,1^+,2^-,3^-,4^-)&=& \bigg\{
-\frac{s_{234}^3}{4\la1|p_{H}|2]\la1|p_{H}|4][23][34]}W^{(1)} \nonumber\\&&
-\bigg(\frac{\la2|p_{H}|1]^3}{2s_{134}\la2|p_{H}|3][34][41]} 
	+\frac{\la34\ra^3m_{H}^4}{2s_{134}\la1|p_{H}|2]\la3|p_{H}|2]\la41\ra}
	\bigg)W^{(2)}\nonumber\\&&
+\frac{1}{4s_{124}}\bigg(\frac{\la3|p_{H}|1]^4}{\la3|p_{H}|2]\la3|p_{H}|4][21][41]}+\frac{\la24\ra^4m_{H}^4}{\la12\ra\la14\ra\la2|p_{H}|3]\la4|p_{H}|3]}
	\bigg)W^{(3)}\nonumber\\&&
- \bigg(\sum_{\gamma=\gamma_\pm(p_H,p_1+p_2)}
	\frac{m_\phi^4\A{\kf1}{2}^3\A 34^3}{\gamma(\gamma+m_\phi^2)\A{\kf1}{1}\A{\kf1}{3}\A{\kf1}{4}\A 12} \bigg)\Ftmtri(m_{H}^2,s_{12},s_{34}) \nonumber\\&&
+\left(1-\frac{N_f}{4N_c}\right)\bigg(\frac{\AB3{p_H}1^2}{s_{124}\B24^2} \FFom(s_{12},s_{14};s_{124})\nonumber\\&&
-\frac{4\la24\ra\la3|p_H|1]^2}{s_{124}[42]}\hat{L}_1\left(s_{124},s_{12}\right)
+\frac{4\la23\ra\la4|p_H|1]^2}{s_{123}[32]}\hat{L}_1\left(s_{123},s_{12}\right)
\bigg)\nonumber\\&&
-\bigg(1-\frac{N_f}{N_c}+\frac{N_s}{N_c}\bigg)\bigg(\frac{\B12\B41\AB3{p_H}2\AB3{p_H}4}{2s_{124}\B24^4}\FFom(s_{12},s_{14};s_{124})\nonumber\\&&
+\frac{2s_{124}\la24\ra\la34\ra^2[41]^2}{3[42]}\hat{L}_3\left(s_{124},s_{12}\right)\nonumber\\&&+\frac{\la34\ra[41]\left(3s_{124}\la34\ra[41]+\la24\ra\la3|p_H|1][42]\right)}{3[42]^2}\hat{L}_2\left(s_{124},s_{12}\right)\nonumber\\&&
+\left(\frac{2s_{124}\la34\ra^2[41]^2}{\la24\ra[42]^3}-\frac{\la24\ra\la3|p_H|1]^2}{3s_{124}[42]}\right)\hat{L}_1\left(s_{124},s_{12}\right)\nonumber\\&&
+\frac{\la3|p_H|1]
(4s_{124}\la34\ra[41]+\la3|p_H|1](2s_{14}+s_{24}))}{s_{124}\la24\ra[42]^3}\hat{L}_0\left(s_{124},s_{12}\right)\nonumber\\&&
-\frac{2s_{123}\la23\ra\la34\ra^2[31]^2}{3[32]}\hat{L}_3\left(s_{123},s_{12}\right)
+\frac{\la23\ra\la34\ra[31]\la4|p_H|1]}{3[32]}\hat{L}_2\left(s_{123},s_{12}\right)\nonumber\\&&
+\frac{\la23\ra\la4|p_H|1]^2}{3s_{123}[32]}\hat{L}_1\left(s_{123},s_{12}\right)
\bigg)\bigg\}\nonumber\\&&
+\bigg\{(2 \leftrightarrow 4)\bigg\}.
\end{eqnarray}
For convenience we have introduced the following combinations of the finite pieces of one-mass $(\FFom)$ and two-mass hard $(\FFtmh)$ box functions (see Appendix~\ref{sec:basint}), 
\begin{eqnarray}
W^{(1)}&=&\FFom(s_{23},s_{34};s_{234})+\FFtmh(s_{41},s_{234};m_H^2,s_{23})+\FFtmh(s_{12},s_{234};s_{34},m_H^2)\nonumber \\
W^{(2)}&=&\FFom(s_{14},s_{34};s_{134})+\FFtmh(s_{12},s_{134};m_H^2,s_{34})+\FFtmh(s_{23},s_{134};s_{14},m_H^2)\nonumber \\
W^{(3)}&=&\FFom(s_{12},s_{14};s_{124})+\FFtmh(s_{23},s_{124};m_H^2,s_{14})+\FFtmh(s_{34},s_{124};s_{12},m_H^2)\nonumber .
\end{eqnarray}
In addition, to simplify the coefficients of the three-mass triangle $\Ftmtri(K_1^2,K_2^2,K_3^2)$ with three off-shell legs $K_1^2,~K_2^2,~K_3^2\neq 0$, we use the notation of eq.~\eqref{eq:flatdef}.
The rational part of the Higgs NMHV amplitude is given by eq.~\eqref{eq:Hrat} (which incorporates the rational $A^{(1)}_4(\phi^{\dagger},1^+,2^-,3^-,4^-)$  amplitude derived in~\cite{Berger:phinite}),
\begin{eqnarray}
R_4(H,1^+,2^-,3^-,4^-)=&&\bigg\{\bigg(1-\frac{N_f}{N_c}+\frac{N_s}{N_c}\bigg)\frac{1}{2}\bigg(\frac{\la 23\ra\la34\ra\la4|p_{H}|1][31]}{3s_{123}\la12\ra[21][32]}
-\frac{\la3|p_{H}|1]^2}{s_{124}[42]^2}\nonumber\\&&+\frac{\la24\ra\la34\ra\la3|p_{H}|1][41]}{3s_{124}s_{12}[42]}
-\frac{[12]^2\la23\ra^2}{s_{14}[42]^2}-\frac{\la24\ra(s_{23}s_{24}+s_{23}s_{34}+s_{24}s_{34})}{3\la12\ra\la14\ra[23][34][42]}\nonumber\\&&
+\frac{\la2|p_{H}|1]\la4|p_{H}|1]}{3s_{234}[23][34]}-\frac{2[12]\la23\ra[31]^2}{3[23]^2[41][34]}  \bigg) \bigg\} + \bigg\{(2\leftrightarrow 4)\bigg\}.
\end{eqnarray}

\section{Numerical Evaluation}
	
\label{sec:num}
In this section we provide numerical values for the helicity amplitudes given in the previous
section at a particular phase space point. To this end, we redefine the finite part of the Higgs amplitude as:
\begin{align}
\lefteqn{	A^{(1)}_4(H,1^{\lambda_1},2^{\lambda_2},3^{\lambda_3},4^{\lambda_4}) 
	= c_\Gamma A^{(0)}(H,1^{\lambda_1},2^{\lambda_2},3^{\lambda_3},4^{\lambda_4})\Bigg( -\frac{1}{\e^2}\sum_{i=1}^4 \left( \frac{-\mu^2}{s_{i,i+1}} \right)^\e} \\&&
	+M^{\mathcal{F},g}_4(\lambda_1,\lambda_2,\lambda_3,\lambda_4) \nonumber
	+\frac{N_f}{N_c}M^{\mathcal{F},f}_4(\lambda_1,\lambda_2,\lambda_3,\lambda_4) 
	+\frac{N_s}{N_c}M^{\mathcal{F},s}_4(\lambda_1,\lambda_2,\lambda_3,\lambda_4) 
	\Bigg).
\end{align}
We evaluate the amplitudes at the phase space point used by Ellis et al. \cite{Ellis:1lh24},
\begin{eqnarray}
p_H^\mu &=& (-1.00000000000, 0.00000000000, 0.00000000000, 0.00000000000),\nonumber \\
p_1^\mu &=& (+0.30674037867, -0.17738694693, -0.01664472021, -0.24969277974) ,\nonumber \\
p_2^\mu &=& (+0.34445032281, +0.14635282800, -0.10707762397, +0.29285022975) ,\nonumber \\
p_3^\mu &=& (+0.22091667641, +0.08911915938, +0.19733901856, +0.04380941793) ,\nonumber \\
p_4^\mu &=& (+0.12789262211, -0.05808504045, -0.07361667438, -0.08696686795) .	
\label{eq:phase}
\end{eqnarray}
The results are presented in table~\ref{tab:fullnum} where we have chosen the renormalisation scale
to be $\mu^2=m_H^2$.\footnote{We have been informed by John Campbell, that the entries for $M_4^{F,g}$ and $M_4^{F,q}$ in Table~\ref{tab:fullnum} are in agreement with results obtained using the seminumerical code described in Ref.~\cite{Ellis:1lh24}.}

\begin{table}[h!]
{\footnotesize
\begin{tabular}[h]{|c|c|c|c|c|}
	\hline
	Helicity & $A^{(0)}$ & $M^{\mathcal{F},g}_4$ & $M^{\mathcal{F},f}_4$ & $M^{\mathcal{F},s}_4$ \\
	\hline
$----$ &  -116.526220-18.681775\,I & -9.540396-0.001010\,I & -0.176850+0.001010\,I & 0.176850-0.001010\,I \\
\hline
$+---$ &  10.308088-0.824204\,I & -10.809925+0.056646\,I & -0.388288+0.198369\,I & 0.296783-0.155132\,I \\
\hline
$--++$ &  20.511457-0.888525\,I & -10.991033+0.320009\,I & 0.268501-0.068414\,I & 0.066595-0.015451\,I \\
\hline
$-+-+$ &  4.683784+4.242678\,I & -10.332320+0.149216\,I & 0.028668-0.066437\,I & 0.166800+0.038844\,I \\
\hline

\end{tabular}
}
\caption{Numerical values for the finite parts of the Higgs + 4 gluon helicity amplitudes at the phase space point given in eq.~(6.2).}
\label{tab:fullnum}
\end{table}

\section{Conclusions}
\label{sec:conc}

We have calculated the last (analytically) unknown building block of the Higgs plus four gluon amplitude, the $\phi$-NMHV amplitude $A^{(1)}_4(\phi,1^+,2^-,3^-,4^-)$. We chose to split the calculation into two parts, one being cut-constructible (to which we applied the techniques of four-dimensional unitarity) and a rational part, which is insensitive to four-dimensional cuts.  

By employing a four-dimensional unitarity-based strategy, we reconstructed the cut-constructible piece
as a combination of 
bubble-, triangle- and box-functions.
The coefficients of the box-integrals were obtained using quadruple cuts \cite{Britto:2004nc}.
We used Forde's Laurent expansion method \cite{Forde:intcoeffs} to derive
the coefficients of the one- and two-mass triangles. The only role of these
functions in the amplitude is to ensure the correct infrared $\epsilon$ pole structure, and as such they do not appear explicitly in our result. 
The coefficients of the three-mass triangle, and the bubble-functions were calculated using two independent techniques; 
the Laurent expansion method, \cite{Forde:intcoeffs} and the triple-cut spinor-integration technique \cite{Mastrolia:2006ki}. 
Finally the double cuts were computed 
analytically  {\it via} Stokes' theorem~\cite{Mastrolia:2009dr} and checked numerically with the Laurent expansion technique~\cite{Forde:intcoeffs}.  

To evaluate the remaining rational piece of the amplitude, we found it convenient to separate the rational part into two terms, one which is sensitive to the number of active light fermions $N_f$
and a homogeneous part which is not. The $N_f$ dependent contribution was efficiently computed using Feynman diagrams, while  
the homogeneous piece was amenable to a BCFW recursion relation approach~\cite{Britto:rec,Britto:proof}.

Previous calculations of $\phi$ plus gluon amplitudes have considered $\phi$-MHV, \cite{Badger:1lphiMHV,Glover:1lphiMHVgeneral}, the $\phi$-all-minus, \cite{Badger:1lhiggsallm} and the $\phi$-all-plus and $\phi$-nearly-all-plus rational amplitudes~\cite{Berger:phinite}.  We have collected the results of these previous papers, so that the virtual one-loop amplitudes for the $ 0 \rightarrow Hgggg$ process (in the large-$m_t$ limit) are available in one place.
In particular, we present compact formulae for the helicity configurations $A^{(1)}_4(H,1^-,2^-,3^-,4^-)$, $A^{(1)}_4(H,1^-,2^-,3^+,4^+)$, $A^{(1)}_4(H,1^-,2^+,3^-,4^+)$
and $A^{(1)}_4(H,1^+,2^-,3^-,4^-)$ in section~\ref{sec:fullres},  knowing that all other helicity configurations can be obtained by parity, or by permuting the gluon labels.

The Higgs plus four gluon amplitudes have been calculated numerically in Ref.~\cite{Ellis:1lh24}. However,  
we envisage that the analytic formulae hereby presented can be used to provide 
a faster and more flexible analysis of Higgs phenomenology at the LHC.  

\acknowledgments 

We wish to thank Keith Ellis and John Campbell for useful discussions and 
important cross checks.  We thank Lance Dixon for a careful reading of the manuscript and for pointing out a number of typographical errors that have now been corrected. SB acknowledges support from the Helmholtz Gemeinschaft
under contract VH-NG-105. EWNG gratefully acknowledges the support of the
Wolfson Foundation and the Royal Society.  PM wishes to thank the IPPP in Durham
for kind hospitality at several stages  while this work was being performed. CW
acknowledges the award of an STFC studentship. This research was supported in
part by the UK Science and Technology Facilities Council and by the European
Commission's Marie-Curie Research Training Network under contract
MRTN-CT-2006-035505 ``Tools and Precision Calculations for Physics Discoveries
at Colliders".

\appendix
\section{Tree-level amplitudes}
\label{sec:tree}

Compact analytic forms for the tree-level amplitudes relevant for
$A_4^{(1)}(\phi,1^+,2^-,3^-,4^-)$ have been computed using both MHV rules \cite{Dixon:HiggsMHV} and BCFW recursion
\cite{Berger:phinite}.

The first main ingredients are the four simple, all-multiplicity results:
\begin{align}
	A^{(0)}_n(\phi,1^+,\ldots,n^+) &= 0,\\
	A^{(0)}_n(\phi,1^+,\ldots,i^-,\ldots,n^+) &= 0,\\
	A^{(0)}_n(\phi,1^+,\ldots,i^-,\ldots,j^-,\ldots,n^+) &= \frac{\A ij^4}{\prod_{k=1}^n \A{k}{k+1} },\\
	A^{(0)}_n(\phi,1^-,\ldots,n^-) &= \frac{(-1)^nm_\phi^4}{\prod_{k=1}^n \B{k}{k+1} }.
\end{align}
In addition we need the non-trivial NMHV-type amplitude,
\begin{align}
	&A^{(0)}_n(\phi,1^+,2^-,3^-,4^-) =\nonumber\\ 
&-\frac{ m_\phi^4 \la 24\ra ^4}{s_{124} \la 12\ra  \la 14\ra \la 2|p_\phi|3] \la 4|p_\phi|3]}
+\frac{ \la 4|p_\phi|1]^3}{s_{123} \la 4|p_\phi|3] [12] [23]}
-\frac{ \la 2|p_\phi|1]^3}{s_{134} \la 2|p_\phi|3] [14] [34]}.
\end{align}

The amplitudes for the $\phi$ coupling to a quark pair and a pair of massless
scalars are given by,
\begin{align}
	A^{(0)}_n(\phi,1_s,\ldots,i^-,\ldots,n_s) &= \frac{\A 1i^2\A ni^2}{\prod_{k=1}^n \A{k}{k+1} },\\
	A^{(0)}_n(\phi,1_q^-,\ldots,i^-,\ldots,n_\qb^+) &= \frac{\A 1i^3\A ni}{\prod_{k=1}^n \A{k}{k+1} },\\
	A^{(0)}_n(\phi,1_q^+,\ldots,i^-,\ldots,n_\qb^-) &= \frac{(-1)^n\A 1i\A ni^3}{\prod_{k=1}^n \A{k}{k+1} }, 
\end{align}
with the non-trivial NMHV amplitudes given by,
\begin{align}
	&A^{(0)}_n(\phi,1_q^+,2^-,3^-,4_\qb^-) =\nonumber\\
	&\frac{\la24\ra^3m_\phi^4}{s_{124}\la14\ra\la2|p_\phi|3]\la4|p_\phi|3]}
	-\frac{\la4|p_\phi|1]^2}{\la4|p_\phi|3][12][23]}
	+\frac{\la2|p_\phi|1]^2\la2|p_\phi|4]}{s_{134}\la2|p_\phi|3][14][34]},\\
	&A^{(0)}_n(\phi,1_s,2^-,3^-,4_s) =\nonumber\\ 
	&-\frac{\la12\ra\la24\ra^2m_\phi^4}{s_{124}\la14\ra\la2|p_\phi|3]\la4|p_\phi|3]}
	+\frac{s_{123}\la4|p_\phi|1]}{\la4|p_\phi|3][12][23]}
	-\frac{\la2|p_\phi|1]\la2|p_\phi|4]^2}{s_{134}\la2|p_\phi|3][14][34]}.
\end{align}

\section{Scalar Basis Integrals}
\label{sec:basint}

\begin{figure}[h]
         \begin{center}
                 \psfrag{1m}{$\FFom$}
                 \psfrag{2me}{$\FFtme$}
                 \psfrag{2mh}{$\FFtmh$}
                 \psfrag{P}{$P^2$}
                 \psfrag{Q}{$Q^2$}
                 \psfrag{s}{$s$}
                 \psfrag{t}{$t$}
                 \psfrag{1}{}
                 \psfrag{2}{}
                 \psfrag{3}{}
                 \psfrag{4}{}
                 \includegraphics[width=0.8\textwidth]{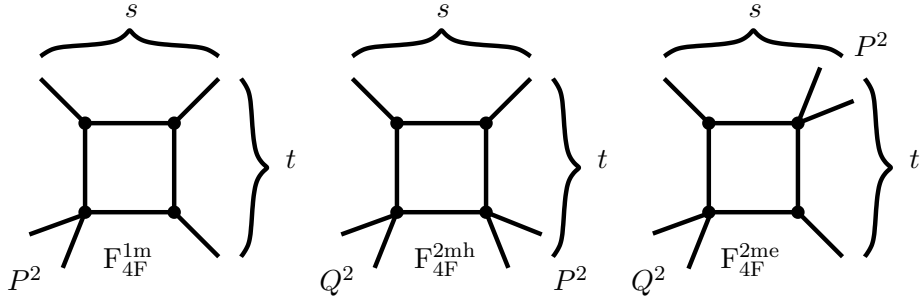}
         \end{center}
         \caption{Conventions for labelling the three scalar box integrals 
appearing in the one-loop $Hgggg$
         amplitudes.}
         \label{fig:boxints}
\end{figure}

In this appendix we present the basis integral functions used to construct the 
various finite contributions to the Higgs helicity amplitudes. Figure~\ref{fig:boxints} sets our labelling conventions.
The finite parts of the one-mass and two-mass (easy and hard) have the following forms,
\begin{eqnarray}
\FFom(s,t;P^2)&=&-2\bigg(\mathrm{Li}_2\bigg(1-\frac{P^2}{s}\bigg)
+\mathrm{Li}_2\bigg(1-\frac{P^2}{t}\bigg)
+\frac{1}{2}\ln^2\bigg(\frac{s}{t}\bigg)+\frac{\pi^2}{6}\bigg),\\
\FFtmh(s,t;P^2,Q^2)&=&-2\bigg(\mathrm{Li}_2\bigg(1-\frac{P^2}{t}\bigg)
+\mathrm{Li}_2\bigg(1-\frac{Q^2}{t}\bigg)
+\frac{1}{2}\ln^2\bigg(\frac{s}{t}\bigg)\nonumber\\
&&-\frac{1}{2}\ln\bigg(\frac{s}{P^2}\bigg)\ln\bigg(\frac{s}{Q^2}\bigg) \bigg),\\
\FFtme(s,t;P^2,Q^2)&=&-2\bigg(\mathrm{Li}_2\bigg(1-\frac{P^2}{s}\bigg)+\mathrm{Li}_2\bigg(1-\frac{P^2}{t}\bigg)+\mathrm{Li}_2\bigg(1-\frac{Q^2}{s}\bigg)\nonumber\\&&+\mathrm{Li}_2\bigg(1-\frac{Q^2}{t}\bigg)-\mathrm{Li}_2\bigg(1-\frac{P^2Q^2}{st}\bigg)+\frac{1}{2}\ln^2\bigg(\frac{s}{t}\bigg)\bigg).
\end{eqnarray}
The finite three-mass triangle is given by~\cite{Bern:pentagon,Lu:3mtri},
\begin{eqnarray}
	\Ftmtri(M_1^2,M_2^2,M_3^2)&=&
	\frac{i}{\sqrt \Delta}
	\sum_{k=1}^3 \left(  \mathrm{Li}_2\left(-\left( \frac{1+i\delta_k}{1-i\delta_k} \right)\right)
	                    -\mathrm{Li}_2\left(-\left( \frac{1-i\delta_k}{1+i\delta_k} \right) \right)\right),
	\label{eq:3mtriINT}
\end{eqnarray}
where,
\begin{eqnarray}
	\Delta &=& \sum_{k=1}^3 -M_k^4+2M_k^2M_{k+1}^2 \\ 
	\delta_k &=& \frac{M_k^2-M_{k+1}^2-M_{k+2}^2}{\sqrt \Delta}.
\end{eqnarray}
Alternative representations for these integrals can be found in references \cite{Binoth:2001vma,vanHameren:1lints,Ellis:2007qk}.

}

\bibliographystyle{JHEP-2}
\providecommand{\href}[2]{#2}\begingroup\raggedright\endgroup
\end{document}